\documentclass[conference]{IEEEtran}
\IEEEoverridecommandlockouts
\usepackage{cite}
\usepackage{amsmath,amssymb,amsfonts}
\usepackage{algorithmic}
\usepackage{graphicx}
\usepackage{textcomp}
\usepackage{xcolor}
\usepackage[T1]{fontenc}
\usepackage{ifpdf}
\usepackage{booktabs}
\usepackage{multirow}
\usepackage[utf8]{inputenc}

\def\BibTeX{{\rm B\kern-.05em{\sc i\kern-.025em b}\kern-.08em
    T\kern-.1667em\lower.7ex\hbox{E}\kern-.125emX}}

\definecolor{refcolor}{rgb}{0,0,0.5}
\ifpdf
\PassOptionsToPackage{hyphens}{url}
\usepackage[%
  pdftitle={Testing pre-trained Transformer models for Lithuanian news clustering},%
  pdfauthor={Lukas Stankevičius, Mantas Lukoševičius},%
  pdfkeywords={},%
  pdfstartview=FitH,%
  bookmarks=true,%
  bookmarksopen=true,%
  breaklinks=true,%
  colorlinks=true,%
  linkcolor=refcolor,%
  anchorcolor=blue,%
  citecolor=refcolor,%
  filecolor=blue,%
  menucolor=blue,%
  urlcolor=refcolor,%
  pdfpagelabels]{hyperref} 
\else 
\usepackage[%
  breaklinks=true,%
  colorlinks=true,%
  linkcolor=blue,%
  anchorcolor=blue,%
  citecolor=blue,%
  filecolor=blue,%
  menucolor=blue,%
  urlcolor=blue]{hyperref} 
\fi 

\begin{document}

\title{Testing
Pre-trained Transformer Models for Lithuanian News Clustering
}

\author{\IEEEauthorblockN{Lukas Stankevičius}
\IEEEauthorblockA{\textit{Faculty of Informatics} \\
\textit{Kaunas University of Technology}\\
Kaunas, Lithuania \\
lukas.stankevicius@ktu.edu}
\and
\IEEEauthorblockN{Mantas Lukoševičius}
\IEEEauthorblockA{\textit{Faculty of Informatics} \\
\textit{Kaunas University of Technology}\\
Kaunas, Lithuania \\
mantas.lukosevicius@ktu.edu}
}

\maketitle

\begin{abstract}
A recent introduction of Transformer deep learning architecture made breakthroughs in various natural language processing tasks. However, non-English languages could not leverage such new opportunities with the English text pre-trained models. This changed with research focusing on multilingual models, where less-spoken languages are the main beneficiaries. We compare pre-trained multilingual BERT, XLM-R, and older learned text representation methods as encodings for the task of Lithuanian news clustering. Our results indicate that publicly available pre-trained multilingual Transformer models can be fine-tuned to surpass word vectors but still score much lower than specially trained doc2vec embeddings. 
\end{abstract}

\begin{IEEEkeywords}
document clustering, document embedding, Lithuanian news articles, Transformer model, BERT, XLM-R, multilingual
\end{IEEEkeywords}

\section{Introduction}

Appearance of a novel Transformer deep learning architecture \cite{vaswani2017attention} sparked a rapid research progress in the Natural Language Processing (NLP) field. Table~\ref{tab1} clearly depicts how quickly models reached human performance on popular NLP evaluation datasets. In less than two years after publication, two evaluation datasets, SQuAD2.0 \cite{rajpurkar2018know} and GLUE \cite{wang2018glue} had human performance outmatched. Currently every top scoring model is of Transformer architecture. The situation was not changed by a newer SuperGLUE \cite{wang2019superglue} task set which had been left with only a tiny gap to human performance. These datasets were among the most popular to evaluate new Transformer models and showed the effectiveness of this new architecture.

\begin{table}[htbp]
\caption{Difficulty of the most popular NLP evaluation datasets}
\begin{center}
\begin{tabular}{l l l l l l}
\toprule
\multicolumn{2}{c}{\textbf{Dataset}} &\multicolumn{4}{c}{\textbf{Score}} \\
\cmidrule(r){1-2} \cmidrule(l){3-6} 
\textbf{Name} & \textbf{Year}& \textbf{Initial} & \textbf{Current}& \textbf{Human} & \textbf{Type} \\
\midrule
RACE \cite{lai2017race} & 2017&44.1&89.4& \textbf{94.5} & Accuracy \\

SQuAD2.0 \cite{rajpurkar2018know}& 2018&66.3 &\textbf{92.58}& 89.542 & F1\\

GLUE \cite{wang2018glue}&2019 &70.0 &\textbf{90.3} & 87.1 & Average \\

SuperGLUE \cite{wang2019superglue} &2019 &71.5 &89.3 & \textbf{89.8} & Average \\
\bottomrule
\end{tabular}
\label{tab1}
\end{center}
\end{table}

There is a need to create NLP models for less-spoken languages. Apart from being less popular than English or Chinese, less-spoken languages also have less content to train the models on. Just the top-10 out of 6\,000 languages in use today make up 76.3\,\% of the total content on the internet\footnote{\url{https://www.internetworldstats.com/stats7.htm}}. Such situation encourages not only to pursue creation of NLP models for other languages but also to look for ways to transfer the knowledge from the content-rich language models.

The most common way to satisfy this need for less spoken languages is to pre-train multilingual models. Examples are LASER \cite{artetxe2019massively} (93 languages), multilingual BERT \cite{devlin2018bert} (104 languages) and XLM-R \cite{conneau2019unsupervised} (100 languages). The authors of XLM \cite{conneau2019cross} showed that training Nepali language model on Wikipedia together with additional data from both English and Hindi decreased perplexity to 109.3 on Nepali, compared to single Nepali training perplexity of 157. Transfer learning and zero-shot translation between language pairs never seen explicitly during training was shown to be possible in \cite{johnson2017google}. Overall, multilingual models can cover many languages, be trained without any cross-lingual supervision and use the bigger languages to benefit the smaller ones.

Lithuanian language does not yet have BERT-scale monolingual NLP model. It is relatively very little spoken in a world. However, as a national language of one of European Union member states, Lithuanian is usually included in the most of pre-trained multilingual models. The aim of this work is to use such Transformer-type models to generate text embeddings and evaluate them on clustering of Lithuanian news articles. Specifically, we will use well known baselines -- multilingual BERT and recently published XLM-R, trained on more than two terabytes of filtered CommonCrawl data.

We chose clustering task to also try to advance the field of data mining. The surge of information, particularly news data, demands tools to help users to ``analyze and digest information and facilitate  decision making'' \cite{aggarwal2012mining}. Unlike classification, clustering is universal in that it can handle unknown categories. Therefore it is well suited for the quickly changing news articles data.

\section{Literature Review}

In this section we review the two first consecutive phases of common natural language processing (NLP)  tasks: text preprocessing and text representation \cite{aggarwal2012mining}. These stages recently are of the most active research and culminated in the development of the Transformer architecture. We also examine relevant NLP contributions for Lithuanian language and our task of news clustering.

\subsection{Text Preprocessing}

Text preprocessing involves selection of features that will bear the understanding of text. The most elementary approach is tokenization into simple characters or words. The finer the tokenization, the smaller is the resulting vocabulary and the more challenging task is given to the NLP model. On the other hand, coarser tokens drastically increase vocabulary size and induce other problems such as sparseness. The middle ground is statistically significant n-grams of both words and chars. Examples of this type of tokenizers are SentencePiece \cite{kudo2018sentencepiece}, BPE \cite{sennrich2015neural}, and WordPiece \cite{wu2016google}. They are often used in the state-of-the-art (SOTA) Transformer models and are shipped together with publicly available pre-trained models. This way manual tokenization step is skipped.

There are number of methods to filter word level tokens. It includes lowercasing, stemming, lemmatization, filtering by maximum and minimum document frequencies (ignoring tokens that are too rare or too common throughout the documents). However, it was shown in \cite{lukas_thesis} that such filtering benefits only the classical text representation approaches such as \textit{tf-idf}, while shallow neural network models, doc2vec \cite{le2014distributed}, benefited of not using any such filtering.

\subsection{Text Representation}

Although tokenized text remains meaningful to us, models still can not operate on it directly. They need it in a numerical form. The preferable way is to derive vector representation for each text sample. Cosine similarity is the simplest example of models operating on (comparing) these embeddings.
The classical approach to text representation uses a Bag of Words (BoW) model. As the name suggests, the order of tokens is lost here and each document is represented by bare counts (histogram) of its tokens. Therefore token weighting such as \textit{tf-idf} is involved. The higher weight of \textit{tf-idf} is, the more descriptive token for a given document is. Given the number of  word $w$ occurrences in a document $d$ as $\textit{tf}_{w,d}$, number of documents containing word $w$ as $\textit{df}_{w}$ and total number of documents $N$, $\textit{tf-idf}_{w,d}$ is given by
\begin{equation}
\textit{tf-idf}_{w,d}=\textit{tf}_{w,d}\cdot \log(N / \textit{df}_w).
\label{eq:tf-idf}
\end{equation}

BoW approaches suffer from several problems. The vector length for each document is the same as the size of the vocabulary. Typically, the vocabulary size is huge and this induces major memory constraints. The embedded vectors are also very sparse as each document uses only the small subset of the vocabulary. Various methods, such as Latent Semantic Analysis (LSA) using Singular Value Decomposition (SVD), are employed to reduce the dimensionality. Nevertheless, SVD has to operate on the same high dimensional $\textit{documents} \times \textit{tokens}$ matrix.

Work of \cite{mikolov2013efficient} revolutionised word embedding calculations. Previous word embeddings, known as co-occurrence vectors, were superseded. They were calculated as direct probabilities of surrounding words in a context window of a given length. The new word2vec \cite{mikolov2013efficient} algorithm uses the same training inputs, except the goal is not to calculate the word distribution, but to derive such embedding weights that context words would be predicted with maximum accuracy. Such setup significantly reduced the word vector size and eliminated problems of high dimensionality and sparseness. Later, the next word embedding model, Global Vectors (GloVe) for word representation was presented \cite{pennington2014glove}. It merged advantages of both the matrix factorization and the shallow window-based methods. Currently it is the most used method for independent word vectors.

Original word2vec word vectors were extended to paragraph vectors by doc2vec model \cite{le2014distributed}. Here each sequence of tokens has its own embedding in the same space as that of words. CBOW and Skip-gram architectures from the original word2vec algorithm were applied to documents correspondingly as PV-DM and PV-DBOW. Distributed Memory model of Paragraph Vector (PV-DM) is tasked to predict the next context word given the previous context and document vector, thus the vector has to sustain memory of that is missing. Distributed Bag of Words version of Paragraph Vectors (PV-DBOW) is forced to predict context vectors randomly, given only the document vector. The original work \cite{le2014distributed} used a concatenation of the both models. However, in \cite{lukas_thesis} it was found that PV-DBOW alone gave the best results.

The most recent text representation models produce contextualised token vectors. Models like ELMO \cite{peters2018deep} and various Transformers have each of their inputs to interact with the other ones. This leads to each token vector being aware of the others. Contextualisation solved the problem of word polysemy seen in word2vec or GloVe models.

The Transformer architecture presented in 2017  \cite{vaswani2017attention} excels other contextualised models due to several reasons. Firstly, only after a single layer each input representation becomes aware of the other ones. For Recurrent Neural Networks (RNN) like ELMO it took $n$ layers, where $n$ is the sequence length. Another advantage over recurrent architectures is that Transformer is very parallelizable. It does not need to wait for a hidden state of the previous word as is the case with RNN. This particular feature led to creation of multi-billion-parameter Transformers such as GPT-2 \cite{radford2019language}, T5 \cite{raffel2019exploring}, Megatron \cite{shoeybi2019megatron}, and T-NLG\footnote{\url{https://www.microsoft.com/en-us/research/blog/turing-nlg-a-17-billion-parameter-language-model-by-microsoft/}}. Despite huge success of Transformer models, it can not process long sequences as its complexity per layer is $O(n^2 \cdot d)$ where $d$ is representation dimension. For example, the maximum input length of the popular BERT model is just 510 tokens. This and other problems of Transformer architecture currently are researched very actively.

\subsection{Related Work on Lithuanian Language}

There are several works on Lithuanian text clustering. \cite{CiganaiteGreta2014Tdc} used internet daily  newspaper  articles  from  the  \textit{Lrytas.lt} news website  and  information  from  the largest  internet  forum  for  mothers -- \textit{supermama.lt}. $k$-means and Expectation Maximization (EM) algorithms were compared on BoW data representation. It was found that optimal clustering algorithm is $k$-means with cosine similarity. Other work analysed unsupervised feature selection for document clustering \cite{VaroneckienAura2014Esou}. Authors found that \textit{tf-idf} weighting with 3\,000 features and spherical $k$-means clustering algorithm works best. A similar observation is expressed in \cite{pranckaitis2017clustering}. Here $k$-means was compared to various hierarchical clustering algorithms and was not superseded. Authors found that \textit{tf-idf} together with stemming is superior to other approaches.

\cite{lukas_thesis} compared BoW and doc2vec (PV-DBOW) Lithuanian news article representations for document clustering. It was shown that PV-DBOW representation, trained on the whole dataset is superior to the BoW method. Authors also investigated various PV-DBOW hyperparameters and outlined recommendations for training method weights.

There is also other NLP work on Lithuanian language. \cite{Jurgita2019SaoL} and \cite{Jurgita2018SaoLsen} compared traditional and deep learning approaches for Lithuanian internet comments sentiment classification. Authors demonstrated that traditional Naïve Bayes Multinomial and Support Vector Machine methods outperformed LSTM and Convolutional neural networks. Other work \cite{urgita2018IeoL} compared CBOW and Skip-gram word embedding architectures and found the first one to be superior. We can add that in this current work we noticed a similar tendency for document vectors: equivalent version of PV-DBOW in our initial experiments outperformed PV-DMM architecture.

We have not found any previous work on Lithuanian language using Transformer models.

\section{The Data}

We followed methodology of \cite{lukas_thesis} and expanded their dataset from 82\,793 up to 260\,146 articles. Although average number of characters in each our text sample is 2\,948, several scraped articles were very small and resulted in empty vectors during averaging the GloVe vectors. Due to this reason we filtered all articles with less than 200 characters and this resulted in a final dataset of 259\,996 texts.

The data consist of Lithuanian news articles scraped from  \textit{lrt.lt}, \textit{15min.lt}, and \textit{delfi.lt} websites. The number of texts are correspondingly 26\,344, 133\,587, and 100\,065. Due to the absence of sitemap in \textit{lrt.lt} website, we did not scrape more articles from this site than is already scraped in \cite{lukas_thesis}.

Evaluation of clustering requires existing knowledge of the potential clusters. For this task we leveraged article category labels extracted from each article URL. Following the mappings of \cite{pranckaitis2017clustering}, the labels were unified from over a hundred categorical descriptions down to 12 distinct categories. The resulting categories of the articles are:
\begin{itemize}
\item Lithuanian news (60\,158 articles);
\item World news (68\,635 articles);
\item Crime (30\,967 articles);
\item Business (19\,964 articles);
\item Cars (6\,313 articles);
\item Sports (14\,910 articles);
\item Technologies (4\,438 articles);
\item Opinions (9\,728 articles);
\item Entertainment (2\,462 articles);
\item Life (3\,811 articles);
\item Culture (7\,967 articles);
\item Other (30\,643 articles that do not fall into the previous categories).
\end{itemize}

During most of experiments we employed a smaller subset of the dataset described above. We sampled randomly 125 news articles from each of the 12 categories. That results in a total of 1\,500 articles equally distributed among the categories and corresponds to the data required for one clustering. We planned to make 50 independent clusterings to average results and enhance their reproducibility. Thus we repeated independent sampling of 1\,500 equally distributed articles and used a subset of 55\,487 unique articles (with repetitions it would be 75\,000). During each experiment we calculated embedding vectors only for those 55\,487 news articles.

\section {Methods}

\subsection{Clustering}

We use $k$-means clustering algorithm. Due to its high speed, it is suitable for large corpora \cite{CiganaiteGreta2014Tdc} and outperforms other clustering algorithms \cite{pranckaitis2017clustering}. During experiments we feed vectorized document representations and the expected number of clusters $k$ to $k$-means and receive document assignments to clusters. We set each of 50 $k$-means initialisations the same.

\subsection{Evaluation}

First, we calculate the following confusion matrix elements:
\begin{itemize}
\item \textit{TP} – pairs of articles that have same category label and are predicted to be in the same cluster;
\item \textit{TN} – pairs of articles that belong to different categories and are predicted to be in different clusters;
\item \textit{FP} – pairs of articles that belong to different categories but are predicted to be in the same cluster;
\item \textit{FN} – pairs of articles that have same category label but are predicted to be in different clusters.
\end{itemize}
We chose to evaluate clusters by Matthews Correlation Coefficient (MCC) score due to its reliability as described in \cite{chicco2020advantages}. It is calculated as
\begin{equation}
\textit{MCC}= \frac{\textit{TP}\cdot\textit{TN} - \textit{FP}\cdot FN}
{\sqrt{(\textit{TP}+\textit{FP})(\textit{TP}+\textit{FN})(\textit{TN}+\textit{FP})(\textit{TN}+\textit{FN})}}.
\label{eq:mcc}
\end{equation}
The MCC score ranges from -1 to 1. Scores around 0 value correspond to random clustering, while close to 1 indicate perfect matching.

\subsection{PV-DBOW}

This doc2vec version was trained on all our dataset -- total of 259\,996 Lithuanian news articles. We preprocessed the dataset by lowercasing and tokenizing it into words. The same vector size (100), number of epochs (10), window (12), and minimum count (4) parameters were used as in \cite{lukas_thesis}. PV-DBOW returns a single embedding for each document so no further aggregation is required.

\subsection{GloVe}

We performed our own text preprocessing during experiments with GloVe \cite{pennington2014glove} type Lithuanian word vectors \cite{lithuanianGlove}. It combined lowercasing and word level tokenization. Out of whole unique 1\,028\,816 tokens from our whole dataset 311\,470 were also present in Lithuanian GloVe vectors. This unique tokens intersection amounts to 30\,\% of our tokens and up to 94\,\% of GloVe.

We tried several ways of aggregating GloVe vectors:
\begin{itemize}
\item calculating an average of all the word vectors in the article;
\item weighting all tokens with \textit{tf-idf} and calculating an average of the 20 word vectors with the highest weight;
\item weighting all tokens in an article with \textit{Softmax(tf-idf)} and calculating a weighted average of all the word vectors in the article.
\end{itemize}

\subsection{Multilingual BERT}

BERT \cite{devlin2018bert} outputs the same number of vectors as it is fed inputs. The first is a special [CLS] token which is designed to be used in sentence level tasks. The following are text data tokens, ending with the last [SEP] token. Optionally, one can add a second [SEP] token in a middle of an input sequence to separate two text segments. In our experiments we input only one segment and always separately try the [CLS] token output vector and the averages of all token vectors.

Pre-trained models like multilingual BERT are supposed to be fine-tuned for the desirable task. We performed Masked Language Modelling fine-tuning on half of our subset data -- 27\,743 news articles. We trained with batch size of 4 for 5 epochs totalling 68\,505 steps for uncased and 75\,985 steps for cased versions of pre-trained multilingual BERT model.

The maximum number of input tokens to BERT is 512, including special tokens. The most articles are within 512 tokens limit but some are longer. We tried to estimate effect of this constraint by trying to feed even fewer tokens and analyzing how mean MCC score changes with the longer input sequence.

We carried out our experiments in Google Colab\footnote{\url{https://colab.research.google.com/}} environment. It offers 12\,GB of RAM and GPU-accelerated  machines which allows an order of magnitude speed up of BERT model compared to CPU.

\subsection{XLM-R}

XLM-R \cite{conneau2019unsupervised} is one of the recent multilingual language models, much bigger than multilingual BERT. It is trained on 2 terabytes of filtered text from which 13.7\,GB is Lithuanian. The huge size of this model limited our experiments. We only calculated outputs of the first 512 tokens for each news article. It took approximately a total of 40 hours.

\section{Results}

\subsection{GloVe}

Results with aggregation of GloVe vectors are presented in Table~\ref{tab2}. It is clearly seen that applying \textit{tf-idf} weighting to select the best tokens to average can significantly surpass the simple average of all vectors.

\begin{table}[htbp]

 \centering
 \caption{Methods of combining GloVe vectors}
\begin{tabular}{ l r r }
\toprule
\multirow{2.6}{*}{\textbf{Vector aggregation}}&\multicolumn{2}{c}{\textbf{MCC score}} \\
\cmidrule{2-3} 
 & \textbf{Mean} & \textbf{Std} \\
\midrule
Average & 0.203 & 0.016 \\

Softmax(tf-idf) weighted average & \textbf{0.264} & 0.017 \\

Average of 20 highest tf-idf tokens & \textbf{0.264} & 0.024 \\
\bottomrule
\end{tabular}
\label{tab2}
\end{table}

\subsection{Multilingual BERT}

\begin{figure*}
\centerline{\includegraphics[width=\textwidth]{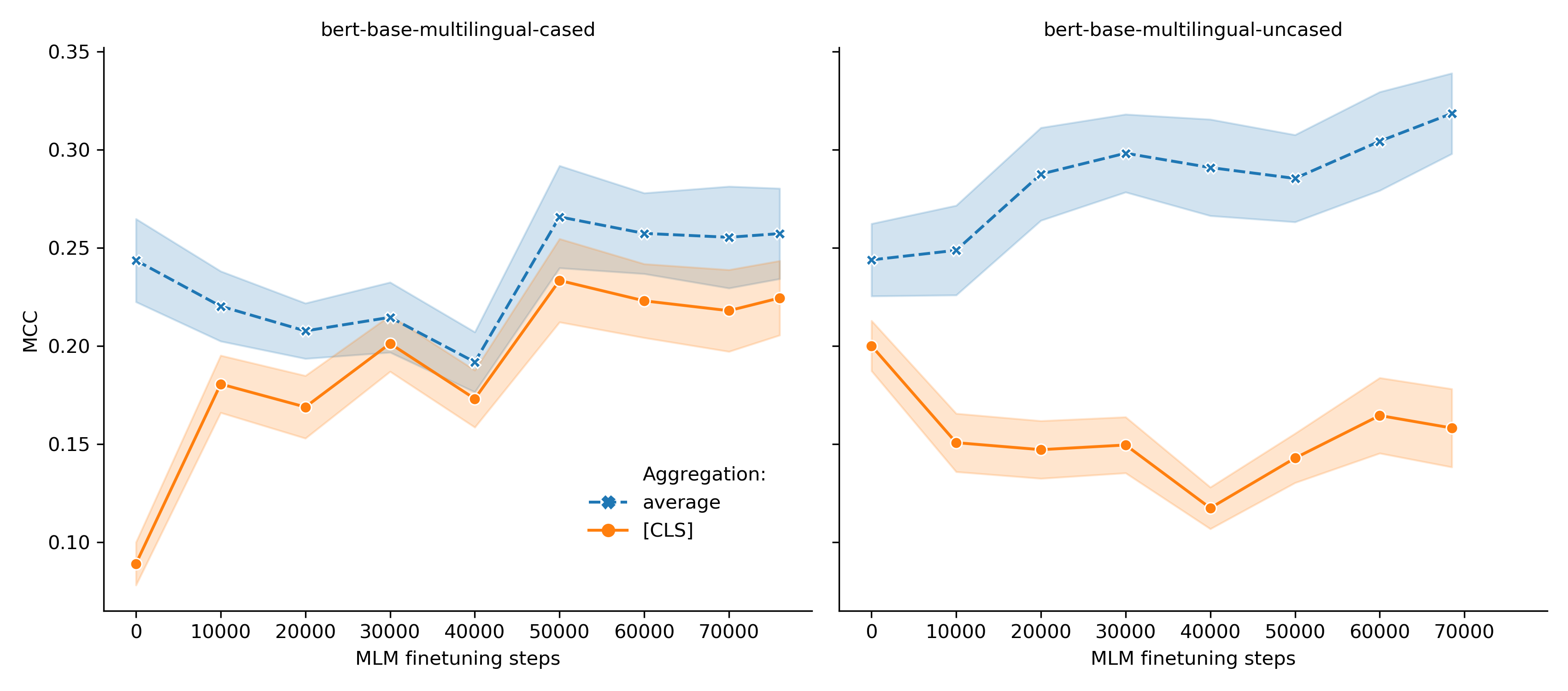}}
\caption{MCC score dependence on multilingual BERT model type (cased or uncased), language modelling fine-tuning steps, and token vector aggregation method (average of all first 128 tokens or just the first [CLS] token). Markers show mean of MCC and shadows -- standard deviation.}
\label{fig}
\centering
\end{figure*}

The effect of multilingual BERT fine-tuning on Lithuanian news articles is depicted in Fig.~\ref{fig}. One can clearly see that (1) the fine-tuning improves the clustering results, (2) the average of all tokens is much better than only the [CLS] vector, and (3) the uncased model version outperforms the cased one.

\begin{figure}
\centerline{\includegraphics[scale=0.6]{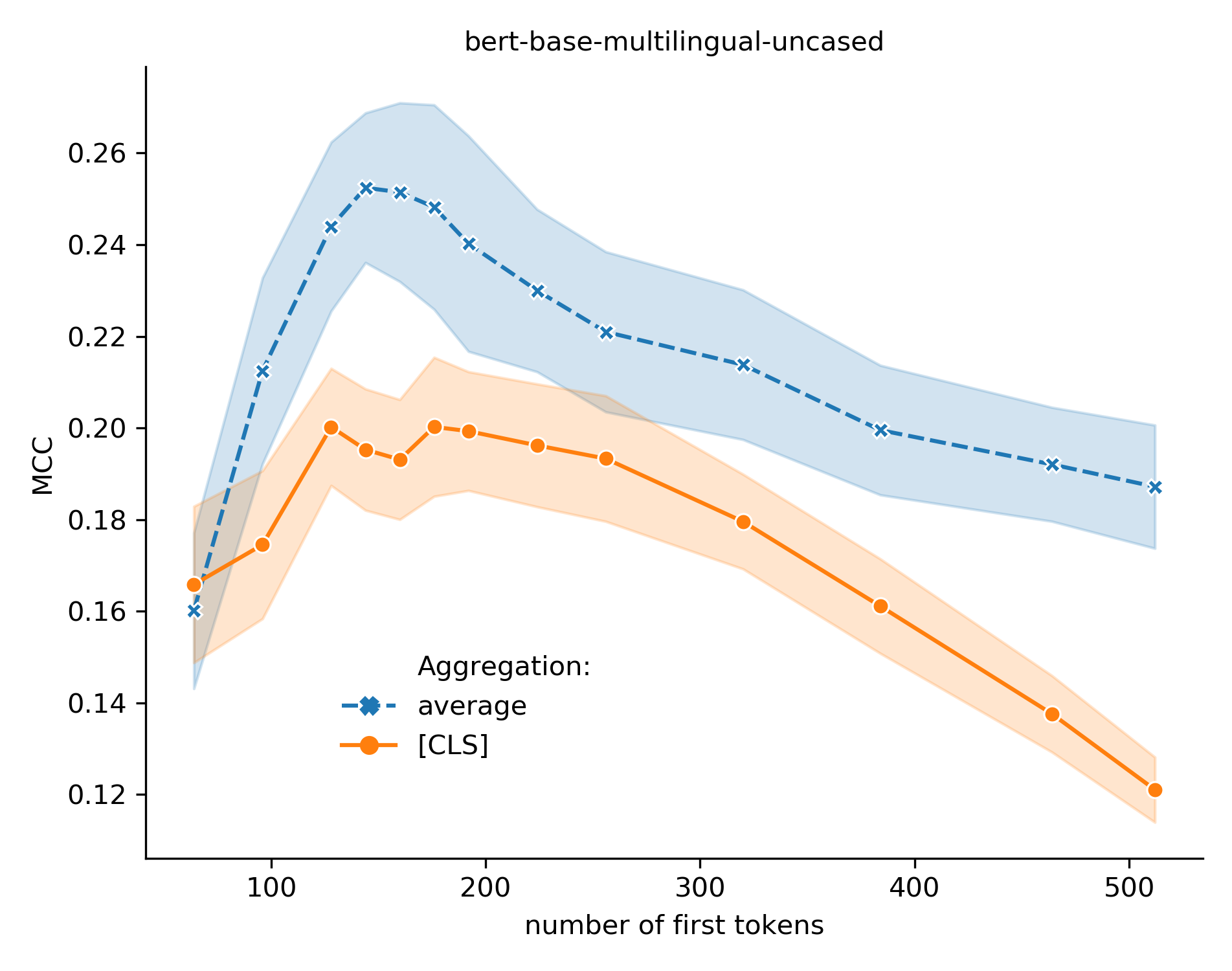}}
\caption{MCC score dependence on the number of first tokens used and the token vector aggregation method (average of all selected tokens or just the first [CLS] token) for multilingual uncased BERT model. Markers show mean of MCC and shadows -- standard deviation.}
\label{fig2}
\centering
\end{figure}

To our surprise, the best results were obtained with limiting the number of tokens to only the first 144 (see Fig.~\ref{fig2}). This can be attributed to the more important information being in the beginning of news article. We observed the same tendency with the XLM-R model.

\subsection{The Best Models}
We tried four different models to represent Lithuanian news articles. Do Transformer models scored better than PV-DBOW? As can be seen in Table~\ref{tab3}, the best Transformer model managed to outperform GloVe vectors. However, PV-DBOW model is far ahead with the mean MCC score of 0.442.

\begin{table}[htbp]
\caption{Comparison of methods}
\begin{center}
\begin{tabular}{ l r r }
\toprule
\multirow{2.6}{*}{\textbf{Text representation method}}&\multicolumn{2}{c}{\textbf{MCC score}} \\
\cmidrule{2-3}
 & \textbf{Mean}& \textbf{Std} \\
\midrule
PV-DBOW& \textbf{0.442} & 0.028 \\
Uncased fine-tuned BERT, average of first 144 tokens & 0.322 & 0.020 \\
Softmax(tf-idf) weighted average of GloVe vectors & 0.264 & 0.017 \\
XLM-R, average of first 288 tokens, total fed 512 & 0.251 & 0.016 \\
\bottomrule
\end{tabular}
\label{tab3}
\end{center}
\end{table}

\section{Conclusions}

In this work we compared multilingual BERT, XLM-R, GloVe, and PV-DBOW text representations for Lithuanian news clustering. For BERT we found out that the average of only the first 144 token vectors outperforms longer aggregations or the [CSL] token vector. We observed that BERT fine-tuning with Lithuanian news articles also improves the results. The other pre-trained Transformer type model XLM-R was too computationally expensive to optimize and out of the four its initial representations scored the worst. Regarding GloVe vectors, we found that its best Softmax(tf-idf) embeddings (mean MCC score 0.264) are outperformed by the BERT. Nevertheless, the best text representation method proved to be PV-DBOW with mean MCC score 0.442. Our work on generating representations for Lithuanian news clustering showed that multilingual pre-trained Transformers can be better than independent GloVe vectors but under-performs against specially trained simpler PV-DBOW.

Multilingual BERT MCC score kept rising till last fine-tuning steps and it is not clear how large the improvement could be accomplished training longer. Our future plan is to clarify this by using more data. We also plan to train a new monolingual BERT model specifically for Lithuanian language. It would be interesting to know if these resource-``hungry'' approaches could surpass the score of the relatively simple PV-DBOW method.

\bibliographystyle{IEEEtran}
\bibliography{text_representations}

\end{document}